\documentclass[twocolumn]{aastex631}
\usepackage{ifthen}
\usepackage{xspace}
\usepackage{upgreek}
\usepackage{nicefrac}
\usepackage{bm}
\usepackage{ulem}
\usepackage{xspace}
\usepackage[italicdiff]{physics}

\definecolor{dg}{rgb}{0.0, 0.6, 0.1}
\definecolor{ed}{rgb}{1.0, 0.6, 0.1}

\newcommand{\Andrew}[1]{\textcolor{dg}{#1}}

\usepackage{pgffor}
\makeatletter
\def\dif{\@ifnextchar[{\@with}{\@without}}
\def\@with[#1]#2{
  \ensuremath{\frac{\foreach \x in {#2}{\mathrm{d}\x\,}}{\foreach \x in {#1}{\mathrm{d}\x\,}}}
}
\def\@without#1{
  \ensuremath{%
    \ifx\hfuzz#1\hfuzz
    \mathrm{d}
    \else
    \foreach \x in {#1}{\mathrm{d}\x\,}
    \fi
    }
}
\makeatother

\newcommand{\unit}[1]{\ensuremath{\,\mathrm{#1}}}

\newcommand{\be}{\begin{equation}}
\newcommand{\ee}{\end{equation}}
\newcommand{\ba}{\begin{eqnarray}}
\newcommand{\ea}{\end{eqnarray}}

\newcommand{\ve}{\ensuremath{\varepsilon}}

\newcommand{\mysub}[1]{\ensuremath{_{\mathrm{#1}}}}
\newcommand{\mysup}[1]{\ensuremath{^{\mathrm{#1}}}}

\newcommand{\myerror}[2][NONE]{%
  \ifthenelse { \equal {#1} {NONE} } %
  {\ensuremath{\pm #2}}%
  {\ensuremath{_{-#1}^{#2}}}%
}

\definecolor{dg}{rgb}{0.0, 0.6, 0.1}
\definecolor{ed}{rgb}{1.0, 0.6, 0.1}
\makeatletter
\def\Andrew{\def\xx{dg}\@ifnextchar[{\@mwith}{\@mwithout}}
\def\Mitya{\def\xx{orange}\@ifnextchar[{\@mwith}{\@mwithout}}
\def\Felix{\def\xx{ed}\@ifnextchar[{\@mwith}{\@mwithout}}
\def\@mwith[#1]#2{\textcolor{\xx}{\sout{#1}#2}}
\def\@mwithout#1{\textcolor{\xx}{#1}}
\makeatother

\newcommand{\dias}{{Dublin Institute for Advanced Studies, School of Cosmic Physics, 31 Fitzwilliam Place, Dublin 2, Ireland}}

\newcommand{\mpik}{{Max-Planck-Institut f\"ur Kernphysik, Saupfercheckweg 1, 69117 Heidelberg, Germany}}
\newcommand{\yerevan}{{Yerevan State University, 1 Alek Manukyan St, Yerevan 0025, Armenia}}
\newcommand{\rikkyo}{{Graduate School of Artificial Intelligence and Science, Rikkyo University, Nishi-Ikebukuro 3-34-1, Toshima-ku, Tokyo 171-8501, Japan}}

\newcommand{\desy}{{DESY, D-15738 Zeuthen, Germany}}

\newif\ifwithmcc

\withmcctrue
\shorttitle{Inverse Compton emission on power-law target photons}
\shortauthors{Khangulyan et al.}
\graphicspath{{./}{figures/}}

\begin{document}

\title{On the properties of inverse Compton spectra generated by up-scattering a power-law distribution of target photons}

\correspondingauthor{Dmitry Khangulyan}
\email{d.khangulyan@rikkyo.ac.jp}

\author[0000-0002-7576-7869]{Dmitry Khangulyan}
\affiliation{\rikkyo}

\author[0000-0003-1157-3915]{Felix Aharonian}
\affiliation{\dias}
\affiliation{\mpik}
\affiliation{\yerevan}

\author[0000-0001-9473-4758]{Andrew M. Taylor}
\affiliation{\desy}



\begin{abstract}
  Relativistic electrons are an essential component in many astrophysical sources, and their radiation may dominate the
  high-energy bands. Inverse Compton (IC) emission is the radiation mechanism that plays the most important role in
  these bands. The basic properties of IC, such as the total and differential cross sections, have long been studied;
  the properties of the IC emission depend strongly not only on the emitting electron distribution but also on the
  properties of the target photons. This complicates the phenomenological studies of sources, where target photons are
  supplied from a broad radiation component. We study the spectral properties of IC emission generated by a power-law
  distribution of electrons on a power-law distribution of target photons. We approximate the resulting spectrum by a
  broken-power-law distribution and show that there can be up to three physically motivated spectral breaks.  If the
    target photon spectrum extends to sufficiently low energies, $\varepsilon_{\mathrm{min}}< m_e^2c^4/E_{\mathrm{max}}$ ($m_e$
    and $c$ are electron mass and speed of light, respectively; $\varepsilon_{\mathrm{min}}$ and $E_{\mathrm{max}}$ are the
    minimum/maximum energies of target photons and electrons, respectively), then the high energy part of the IC component
    has a spectral slope typical for the Thomson regime with an abrupt cutoff close to $E_{\mathrm{max}}$.  The spectra
    typical for the Klein-Nishina regime are formed above $m_e^2c^4/\varepsilon_{\mathrm{min}}$. If the spectrum of target photons features a cooling break, i.e., a change of
    the photon index by $0.5$ at $\varepsilon_{\mathrm{br}}$, then the transition to the Klein-Nishina regime proceeds through an
    intermediate change of the photon index by $0.5$ at $m_e^2c^4/\varepsilon_{\mathrm{br}}$.
\end{abstract}



\section{Introduction}\label{sec:intro}

Inverse Compton (IC) scattering together with synchrotron emission are essential leptonic radiation
mechanisms. Synchrotron--IC models often provide very good fits to broad-band observations for 
astrophysical sources containing ultrarelativistic electrons.  Calculating synchrotron-IC spectral energy distributions (SEDs) is a standard task in
high-energy astrophysics thanks to detailed theoretical descriptions \citep[for a review see][]{1970RvMP...42..237B} and
convenient software packages \citep[e.g., \texttt{naima} by][]{2015ICRC...34..922Z}.
The standard treatment of magnetobremsstrahlung emission is based on a well-known formula from classical electrodynamics that describes the
electromagnetic field of a charge gyrorating in a homogeneous magnetic field, leading to the generation of synchrotron radiation.
If the magnetic field has chaotically distributed directions, the emission spectrum remains essentially unchanged
\citep{1986A&A...164L..16C,2010PhRvD..82d3002A}. This is, however, not a general result. If the magnetic field additionally
features significant fluctuations in its strength, then the standard synchrotron emission spectrum can be considerably
modified \citep{2013ApJ...774...61K,2019ApJ...887..181D}. Furthermore, if the emission is generated within  small-scale
turbulence, in the so-called jitter regime, the produced spectra deviates strongly from that expected in the
conventional synchrotron regime \citep[see in][and references therein]{2013ApJ...774...61K}. Finally, the synchrotron
approximation is not applicable when the emitting particle moves at a small pitch angle to the magnetic
field, such that the curvature of the trajectory of the particle is determined by the curvature of the magnetic field lines
\citep{1996ApJ...463..271C,2015AJ....149...33K}; or when the particle interacts with the magnetic field in the quantum
regime \citep[e.g.,][]{1954PNAS...40..132S}. Despite these physically motivated exceptions, standard synchrotron emission remains an
almost universal approximation for the magnetobremsstrahlung radiation channel.

In the case of IC scattering, the situation is quite different, and the distribution of the target photons plays 
a critical role almost in all astrophysical scenarios. Therefore, to obtain the SED of IC radiation one needs to convolve the differential cross section with the energy and angular distribution of the target photons. The anisotropic differential cross section for IC scattering is available in the literature \citep{1981Ap&SS..79..321A}. It can be readily used for obtaining the IC emission generated by an electron on a monoenergetic beam of photons. If photons have energy and/or angular distribution one can numerically integrate over the photon spectrum or use some analytical derivations available in the literature. These include, for example, the IC cross section averaged over the scattering angle \citep{1968PhRv..167.1159J}, and convolution with a Planckian energy distribution of the target photons
\citep{2014ApJ...783..100K}. 

In what follows, we qualitatively discuss the properties of IC emission generated on a power-law distribution of target
photons.  While numerically computing the corresponding IC spectrum is still a simple task, we focus on finding the key
factors that determine the spectral properties. Our findings simplify the phenomenological analysis of broadband spectra
from gamma-ray sources that feature bright X-ray emission. For example, the results obtained can be used for studying
gamma-ray bursts (GRBs) and flares associated to relativistic outflows from active galactic nuclei (AGN). The
manuscript is organized as follows: in Sect.~\ref{sec:ic} we summarize the properties of IC scattering and introduce
several simple approximations that allow the calculation of the IC loss rates and the mean upscattered photon energy; in
Sec.~\ref{sec:delta} using the \(\delta\)-function approximation {for the single-electron emissivity} we reveal the key factors that determine the spectral
properties of IC component generated on a broad power-law distribution of target photons; and we discuss our
  findings in Sec.~\ref{sec:sum}.

\section{Compton scattering}\label{sec:ic}
The cross section, \(\sigma\mysub{ic}\), describing the scattering of photons by an electron can be obtained with the standard means of quantum electrodynamics. If scattering proceeds in a monodirectional beam of target photons that have energy of \(\ve\), then for a single electron the scattering rate, \(\dot{N}\mysub{ani}\), is given by the usual expression
\be
\dot{N}\mysub{ani} = c\qty(1-v_e\cos\theta) \sigma\mysub{ic}n\mysub{ph}\,.
\ee
Here, \(c\) is speed of light; \(\theta\) is the angle between the beam direction and the electron velocity; and \(v_e\) is electron speed in speed of light units (\(v_e\approx1\) for ultrarelativistic electrons); and $n\mysub{ph}$ is the number of target photons per unit of volume in the laboratory frame. The cross section is given by the following expression \citep[see, e.g.,][]{Landau4}:
\be
\begin{split}
  \sigma\mysub{ic}&=
                    \frac{3\sigma\mysub{T}}{4b_\theta}
                    \left[\qty(1-\frac4{b_\theta}-\frac8{b_\theta^2})\log(1+b_\theta)+\right.\\
  &\left.\frac12+\frac8{b_\theta}-\frac1{2(1+b_\theta)^2}\right]\,,
\end{split}
\ee
where \(b_\theta=2E\ve(1-v_e\cos\theta)/(m_e^2c^4)\) is a parameter that determines the scattering regime; \(\sigma\mysub{T}=8\pi r_e^2/3\) is the Thomson cross section (here $r_e=e^2/(m_{e}c^2)$ is the electron classical radius and \(m_e\) is electron mass). For the sake of simplicity, in what follows we set \(v_e=1\). For \(b_\theta\ll1\) and \(b_\theta\gg1\) the scattering proceeds in the Thomson and Klein-Nishina regimes, respectively.  Since \(b_\theta\) is a Lorentz invariant (indeed, \(b_\theta\propto (pk)\), where \(p\) and \(k\) are four-momenta of interacting electron and photon, respectively), the scattering regime does not depend on the choice of the reference frame.

In any realistic configuration, the target photons have some energy and/or angular distribution: \(\dd{n}\mysub{ph}=n\mysub{ph}\qty(\ve,\bm n)\dd{\ve}\dd{\Omega}_{\bm n}\), where \(\dd{\Omega}_{\bm n}\) is solid angle element in the direction of the photons' momentum, \(\bm n\). To obtain the scattering rate, one needs to integrate over the photon distribution:
\be
\dot{N} = c\int \qty(1-\cos\theta) \sigma\mysub{ic}n\mysub{ph}\qty(\ve,\bm n)\dd{\ve}\dd{\Omega}_{\bm n}\,.
\ee
Here, \(\theta\) is the angle between the electron's initial velocity vector and \(\bm n\). In the relativistic case it is safe to assume that the up-scattered photon propagates in the direction of the electron's initial velocity vector; thus the angular distribution of up-scattered photons is determined both by the electron and target photon distributions. 

If the target photons are isotropically distributed in the laboratory frame: \({n}\mysub{ph}(\ve,\bm n)=n\mysub{ph}(\ve)/(4\pi)\), then the integration over the angular variables can be performed analytically:
\be
\begin{split}
\dot{N}\mysub{iso} &= c\int \dd{\ve}n\mysub{ph}(\ve)\qty[\frac1{4\pi}\int \qty(1-\cos\theta) \sigma\mysub{ic}(b_\theta)\dd{\Omega}_{\bm n}]\,,\\
                   &= c\int \dd{\ve}n\mysub{ph}(\ve)\bar{\sigma}\mysub{ic}(b)\,,
\end{split}
\ee
where \(b=4E\ve/(m_e^2c^4)\) is a dimensionless parameter. We note that the energies \(E\) and \(\ve\) are written in the specific reference frame in which the photon field is isotropic.

The cross section averaged over the scattering angle can be computed analytically yielding a relatively simple expression that, however, contains a dilogarithm function:
\be\label{eq:iso_rate}
\bar{\sigma}\mysub{ic}=
\frac{3\sigma\mysub{T}}{2b^2}\times
\ee
\[
\qty(\frac{b^2+9b+8}{b}\log(b+1) + \frac{2-b^2}{2(b+1)}-9+4\mathrm{Li}_2(-b))\,.
\]
Here \(\mathrm{Li}_2\) is dilogarimth function defined as \(\mathrm{Li}_2(x) = \int\limits_x^0\dd{t}\frac{\log(1-t)}{t}\).
We introduce an auxiliary function \(F\mysub{n,iso}\) by factoring out the dependence on \(b\) as \({\bar{\sigma}\mysub{ic}=\sigma\mysub{T}}F\mysub{n,iso}\qty(b)\). The asymptotic behavior of this function is
\be\label{eq:iso_rate_asym}
\begin{array}{lc}
  F\mysub{n,iso}=\left\{
  \begin{matrix}
    1, & b\ll 1\,\\
    \frac{3}{2b}\log(b), &  b\gg 1\,.
  \end{matrix}
  \right.                                             
\end{array}
\ee
Similar to \cite{2014ApJ...783..100K}, we suggest the following approximate representation for the function $F\mysub{n,iso}$:
\be\label{eq:rate_iso_zero}
G^{(0)}\mysub{n,iso}=\frac{3}{2b}\log\qty(1+\frac{2}{3}b)\,.
\ee
This simple function provides a rough approximation for \(F\mysub{n,iso}\), with a relative error at the level of \(20\%\). If a higher precision is needed, and using the original analytic expression given by Eq.~(\ref{eq:iso_rate}) is not convenient (e.g., because of the presence of dilogarithm function), then the approximation can be improved with the standard correction function from \cite{2014ApJ...783..100K}:
\be\label{eq:correction}
g_{i}\left(x\right)=\left[1+\frac{a_{i}x^{\alpha_{i}}}{1+b_{i}x^{\beta_{i}}}\right]^{-1}\,.
\ee
For example, for the following parameters \(\alpha_i= 0.89\), \(a_i= 0.24\), \(\beta_i= 1.36\), and \(b_i= 0.4\), function 
\be\label{eq:losses_iso}
G\mysub{n,iso}=G^{(0)}\mysub{n,iso}\times g_{i}
\ee
%
approximates the analytical expression for the scattering rate, Eq.~(\ref{eq:iso_rate}), with an accuracy of better than \(0.7\%\).

While the total cross section has a relatively simple mathematical form, obtaining the differential cross section is a more challenging task. The differential cross section, \(\dd{\sigma\mysub{ic}}/\dd{\omega}\), defines the rate of upscattering of target photons with energy \(\ve\) in to the energy interval of \((\omega,\,\omega+\dd{\omega})\).  For astrophysical
applications, the general expressions for the differential cross section can be
significantly simplified using the fact that the energy of the target
photons is typically small, $\ve\ll m_ec^2$, and the electrons are
relativistic, $E\gg m_ec^2$. Under these assumptions, for a monodirectional beam of target photons, the
scattering rate by an electron moving with a velocity that makes an
angle $\theta$ with the photon's direction has the following simple
form \citep{1981Ap&SS..79..321A}:
\be\label{eq:aa_ani}
\dot{n}\mysub{\rm ani}(\omega)=c\left(1-\cos\theta\right)\dv{\sigma\mysub{ic}}{\omega}{n}\mysub{ph}\,,
\ee
where \(\dot{n}\mysub{\rm ani}(\omega)=\dd{\dot{N}\mysub{\rm ani}}/\dd{\omega}\) and
\be
\begin{split}
  \dv{\sigma\mysub{ic}}{\omega}&=
  \frac{3\sigma\mysub{T}}{2b_\theta E}\times\\
&\left[{1+\frac{z^2}{2(1-z)}-\frac{2z}{b_\theta(1-z)}+\frac{2z^2}{b_\theta^2(1-z)^2}}\right]\,.
\end{split}
\ee
Here $z=\omega/E$ is the ratio of the upscattered photon energy to the initial electron energy.  If the target photon field is isotropic, the above expression should be averaged over the interaction angle:
\be\label{eq:aa_iso}
\dot{n}\mysub{\rm iso}(\omega)=c\int\qty(1-\cos\theta)\dv{\sigma\mysub{ic}}{\omega}\,\frac{\dd{\Omega}_{\bm n}}{4\pi}{n}\mysub{ph}=c\bar{\dv{\sigma\mysub{ic}}{\omega}}{n}\mysub{ph}\,.
\ee
Here \(\dot{n}\mysub{\rm iso}(\omega)={\dd{\dot{N}}\mysub{\rm iso}}/{\dd{\omega}}\) and \(\dd{\bar \sigma\mysub{ic}}/\dd{\omega}\) is the angle-averaged cross section \citep{1968PhRv..167.1159J}:
\be
\begin{split}
  \bar{\dv{\sigma\mysub{ic}}{\omega}}&=
                                       \frac{3\sigma\mysub{T}}{bE}\times\left[1+\frac{z^2}{2(1-z)}+\frac{z}{b(1-z)}-\right.\\
  &\left.\frac{2z^2}{b^2(1-z)^2}-\frac{z^3}{2b(1-z)^2}-\frac{2z}{b(1-z)}\,\log\frac{b(1-z)}{ z}\right]\,.
\end{split}
\ee
%
The differential cross section is used to compute the gamma-ray spectrum produced by an electron distribution, \(\dd{N_e}=n_e\dd{E}\), in the anisotropic and isotropic regimes:
\be\label{eq:spectrum_general}
\begin{split}
\dot{n}\mysub{ani}\mysup{tot}(\omega) &=
  c\int  \qty(1-\cos\theta) \dv{\sigma\mysub{ic}}{\omega} n_e(E)n\mysub{ph}(\ve)\dd{E}\dd{\ve}
  \\
  \dot{n}\mysub{iso}\mysup{tot}(\omega) &=  c\int  \dv{\bar{\sigma}\mysub{ic}}{\omega} n_e(E)n\mysub{ph}(\ve)\dd{E}\dd{\ve}
\end{split}
\ee
Another important aspect is that the differential cross section allows one to obtain the IC energy losses of an electron:
\be\label{eq:losses_general}
\begin{split}
\dot{E}\mysub{ani} &=
  -c\int  \qty(1-\cos\theta)\qty(\omega-\ve) \dv{\sigma\mysub{ic}}{\omega} n\mysub{ph}(\ve)\dd{\omega}\dd{\ve}\,,\\
\dot{E}\mysub{iso}&=  -c\int  \qty(\omega-\ve)\dv{\bar{\sigma}\mysub{ic}}{\omega} n\mysub{ph}(\ve)\dd{\omega}\dd{\ve}\,,
\end{split}
\ee
where the integration over \(\omega\) is performed in the range allowed by the kinematic constraints:
\be
\omega\mysub{max/min}=\omega\mysub{cms}\Gamma\mysub{cms}\qty(1\pm v\mysub{cms})\,.
\ee
Here, quantities with subscript ``cms'' are evaluated in the center-of-mass (CMS) reference frame:
\be
\begin{matrix}
  \omega\mysub{cms}&=&\frac12\frac{m_ec^2b_\theta}{\sqrt{1+b_\theta}}\,,\\
  \Gamma\mysub{cms}&=&\frac{E+\omega}{m_ec^2\sqrt{1+b_\theta}}\,,\\
  v\mysub{cms}&=&\sqrt{1-\Gamma\mysub{cms}^{-2}}\,.\\
\end{matrix}
\ee
For relativistic electrons in Eq.~\eqref{eq:losses_general} it is safe to use the following approximations
\be\label{eq:kinematics}
\begin{matrix}
  \omega\mysub{min}&\approx&0\,,\\
  \omega\mysub{max}&\approx&E\frac{b_\theta}{1+b_\theta}\,,\\
  \omega-\ve&\approx&\omega\,.
\end{matrix}
\ee
Note that if an electron interacts with an isotropic photon field, then in Eq.~\eqref{eq:kinematics} one should replace \(b_\theta\) with \(b\) as the upscattered photon energy is maximal for \(\theta=\pi\).

\begin{figure}
  \includegraphics[scale=0.2]{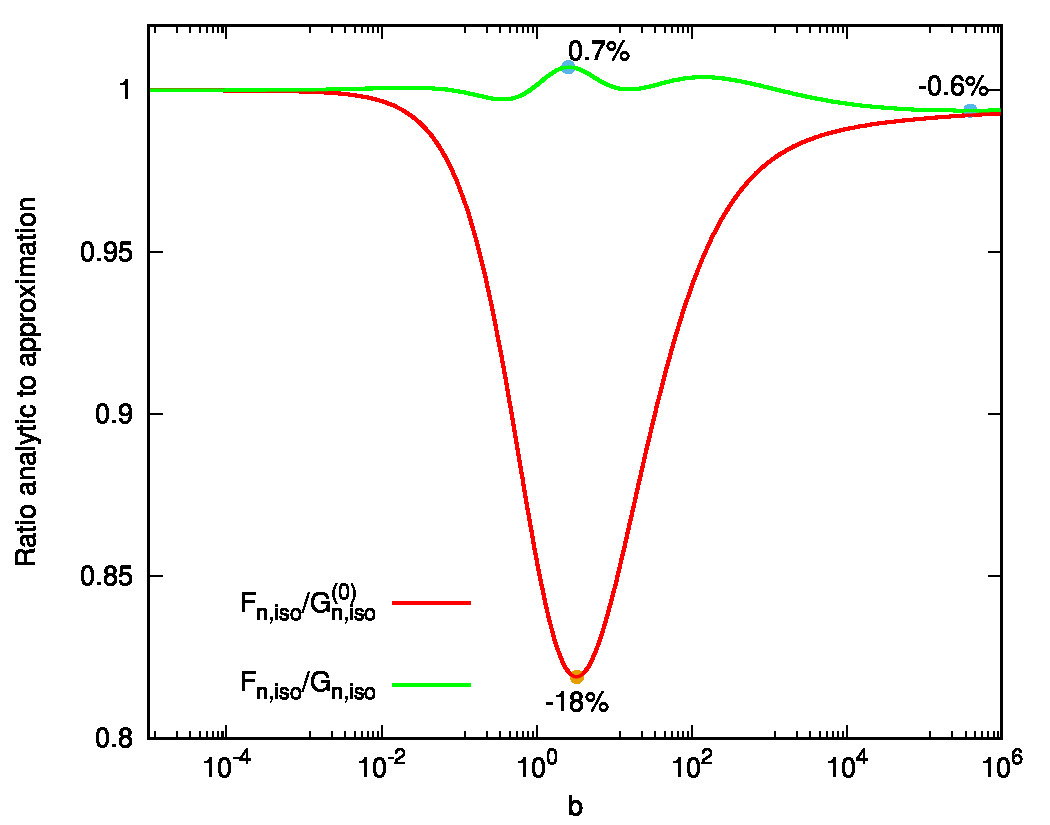}
  \includegraphics[scale=0.1]{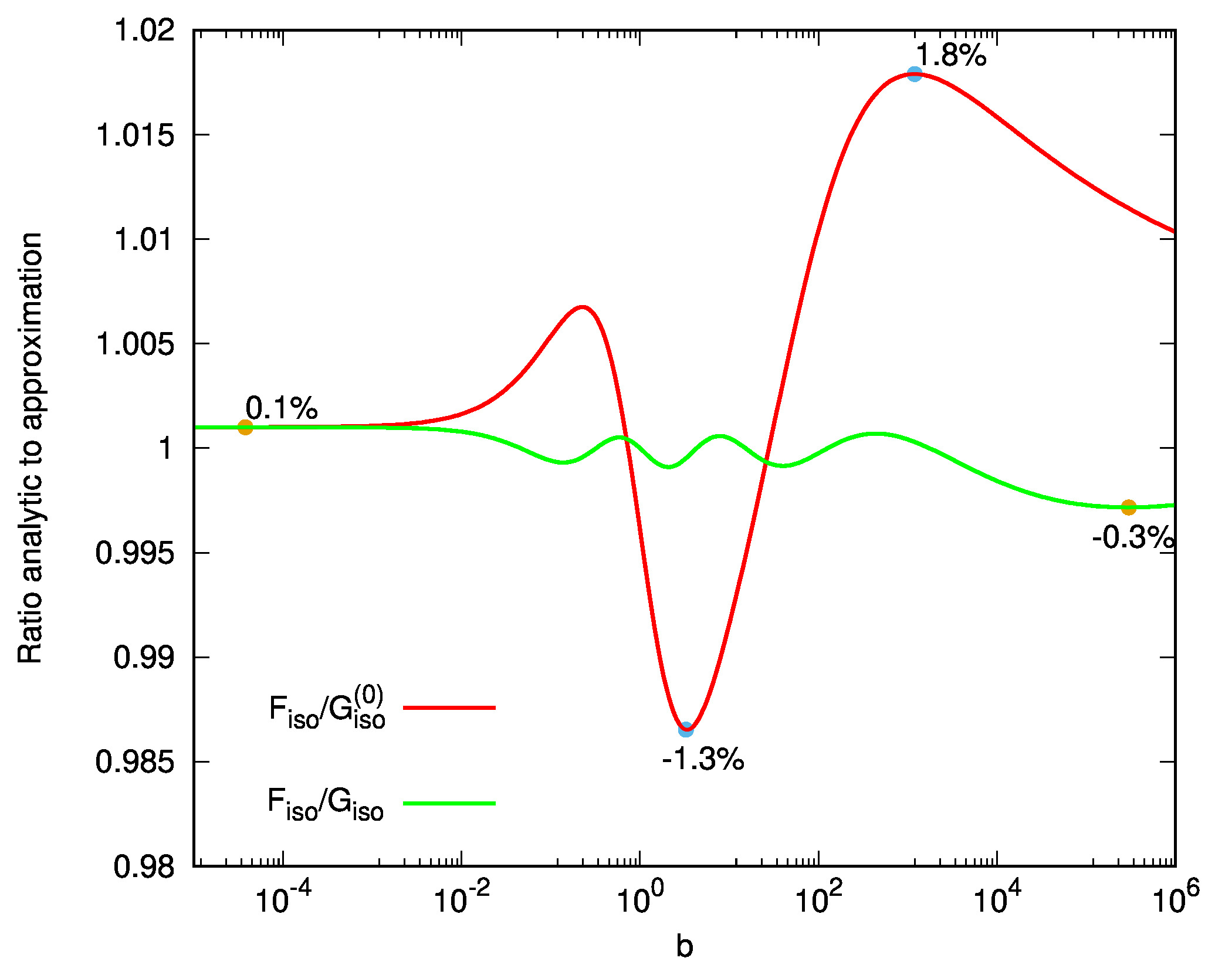}
  \caption{Top panel: The ratio of the function $F\mysub{n,iso}$ to $G\mysub{n,iso}^{(0)}$  and  $F\mysub{n,iso}$ to $G\mysub{n,iso}^{(0)}\times g_i$(for \(a_{i}=0.24\), \(\alpha_{i}=0.89\), \(b_{i}=0.4\), and \(\beta_{i}=1.36\)). Bottom panel: The ratio of the function $F\mysub{iso}$ to $G\mysub{iso}^{(0)}$ (for $c\mysub{iso}=0.785$) and  $F\mysub{iso}$ to $G_{\rm iso}^{(0)}\times g_i$(for $c\mysub{iso}=0.87$, \(a_{i}=-0.275\), \(\alpha_{i}=1.02\), \(b_{i}=4.24\), and \(\beta_{i}=1.1\)). \label{fig:iso}}
\end{figure}

IC energy losses on a monoenergetic beam of photons, \(n\mysub{ph}(\ve)=n_0\delta\qty(\ve-\ve_0)\), can be obtained from Eq.~\eqref{eq:losses_general} by an elementary integration:
\be
\dot{E}\mysub{ani}=-\frac{3\sigma\mysub{T}cn_0E\qty(1-\cos\theta)}{2b_\theta^2}\times
\ee
\[
  \left(\qty(\frac{b_\theta}{2}-2-\frac{6}{b_\theta})\log(1+b_\theta)+\frac{3b_\theta^2+11b_\theta+6}{12(b_\theta+1)^3}+\right.
  \]
  \[
    \left.\frac{11}{2}-\frac{5}{12}b_\theta\right)\,,
\]
where \(b_\theta=2E\ve_0\qty(1-\cos\theta)/(m_e^2c^4)\). 
According to \cite{1968PhRv..167.1159J}, IC energy losses on a mono-energetic isotropic distribution of photons is
\be\label{eq:losses_jones}
\begin{split}
  \dot{E}\mysub{iso}&=-\frac{3\sigma\mysub{T}cn_0E}{b^2}\times\left\{\qty(\frac12b+6+\frac6b)\log\qty(1+b)-\right.\\
  &\left.\frac{\qty[\frac{11}{12}b^3+6b^2+9b+4]}{\qty(1+b)^2}-2+2\mathrm{Li}_2\qty(-b)\right\}\,,\\
         &=-{\sigma\mysub{T}cn_0E}F\mysub{iso}(b)\,,
\end{split}
\ee
where \(b=4\ve_0 E/(m_e^2c^4)\). 
The asymptotic behavior of \(F\mysub{iso}\) is
\be\label{eq:iso_losse_asym}
\begin{array}{lc}
  F\mysub{iso}=\left\{
  \begin{matrix}
    \frac{b}{3},&  b\ll 1\,\\
    \frac{3}{2b}\log(b),&  b\gg 1\,.
  \end{matrix}
  \right.                                             
\end{array}
\ee
Similarly to \cite{2014ApJ...783..100K}, we suggest the following approximate representation for the function  $F_{\rm iso}$
\be\label{eq:losses_iso_zero}
G^{(0)}\mysub{iso}=\frac{3c\mysub{iso}\log(1+0.111b/c\mysub{iso})}{1+2c\mysub{iso}\,b}\,.
\ee
Here \(c\mysub{iso}\) is a numerical factor, which does not change the asymptotic behavior. For example for \(c\mysub{iso}=0.785\), function \(G^{(0)}\mysub{iso}\) follows function \(F\mysub{iso}\) within \(2\%\) margin. This very simple approximation likely provides an accuracy sufficient for any astrophysical application. If a higher precision is needed, one can use the original analytic expression given by Eq.~(\ref{eq:losses_jones}) or improve the approximation with the correction function Eq.~\eqref{eq:correction}. For example, for the following parameters \(c\mysub{iso}=0.87\), \(a_{i}=-0.275\), \(\alpha_{i}=1.02\), \(b_{i}=4.24\), \(\beta_{i}=1.1\), function 
\be
G\mysub{iso}=G^{(0)}\mysub{iso}\times g_{i}
\ee
approximates the analytic expression for IC losses, Eq.~(\ref{eq:losses_jones}), with accuracy better than \(0.3\%\). Comparison of approximations with the analytic expression is shown in Fig.~\ref{fig:iso}.

The mean photon energy can be obtained from the energy loss and scattering rates:
\be\label{eq:mean_photon}
\bar{\omega}\mysub{iso}\qty(E,\ve) = \frac{\dot{E}\mysub{iso}}{\dot{N}\mysub{iso}}=E\frac{F\mysub{iso}(b)}{F\mysub{n,iso}(b)}\,.
\ee
In the Thomson and Klein-Nishina regimes, this yields the well-known asymptotic expressions
\be\label{eq:mean_photon_iso_asym}
\frac{\bar{\omega}\mysub{iso}}{E}=\left\{
  \begin{matrix}
    b/3&\qq{if}& b\ll1\,,\\
    1&\qq{if}&b\gg1\,.
  \end{matrix}
\right.
\ee

To express the mean photon energy in the transition regions, Eq.~\eqref{eq:mean_photon} can be approximated as
\be
\frac{\bar{\omega}\mysub{iso}\qty(E,\ve)}{E} \approx \frac{G\mysub{iso}^{(0)}(b)}{G\mysub{n,iso}^{(0)}(b)}\,,
\ee
which can be further reduced to the simple expression,
\be
\frac{\bar{\omega}\mysub{iso}\qty(E,\ve)}{E} \approx \frac{6b/5}{1+6b/5}\frac{\log\qty(1+5b/27)}{\log\qty(1+2b/3)}.
\ee
This provides a better than \(2\%\) approximation for the upscattered photon energy in the entire energy range.

A similar expression is available for the anisotropic scattering regime. In this case, the asymptotic behavior is
\be\label{eq:mean_photon_ani_asym}
\frac{\bar{\omega}\mysub{ani}}{E}=\left\{
  \begin{matrix}
    b_\theta/2&\qq{if}& b_\theta\ll1\,,\\
    1&\qq{if}&b_\theta\gg1\,.
  \end{matrix}
\right.
\ee
If the scattering proceeds in the anisotropic regime, the following expression provides a better than \(2\%\) approximation for the upscattered photon energy in the entire energy range
\be
\frac{\bar{\omega}\mysub{ani}\qty(E,\ve)}{E} \approx \frac{2b_\theta}{1+2b_\theta}\frac{\log\qty(1+b_\theta/2)}{\log\qty(1+2b_\theta)}\,.
\ee

\section{$\delta$-function approximation}\label{sec:delta}
Both in the classical Thomson and quantum Klein-Nishina regimes, a monoenergetic distribution of electrons generates a broad IC component. If the electrons themselves feature a spread in their energy distribution, then the IC component is further broadened. Once the relative width of the electron distribution exceeds the relative width of single-electron IC spectrum, the width of the single-electron IC spectrum has little influence on the total IC component, and can thus be neglected. This can be appreciated by considering IC scattering under the \(\delta\)-function approximation. In this \(\delta\)-approximation treatment, the electron energy loss rate, \(\dot{E}\), and mean frequency of IC photons generated by this electron, \(\bar{\omega}\), determine the emission spectrum generated by the electron \citep[see, e.g.,][]{1966ApJ...146..686F,2005ICRC....4..131K}:
\be\label{eq:d-function}
\dot{n}_{\gamma,0}=\abs{\frac{\dot{E}}{\omega}}\delta\qty(\omega-\bar{\omega})\,,
\ee
where \(\dot{n}_{\gamma,0}\) is the number of photons upscattered per unit time and frequency:
\be
\dot{n}_{\gamma,0}=\dv{\dot{N}_{\gamma,0}}{\omega}\,.
\ee
Equation~(\ref{eq:d-function}) reproduces correctly the scattering rate and the radiation energy losses of the electrons emitted.
The spectrum produced by an ensemble of electrons is obtained by convolution:
\be\label{eq:d-spectrum}
\dot{n}_{\gamma}=\int \dd{E} \dot{n}_{\gamma,0} n_e\,.
\ee
Here \(n_e\) is electron energy distribution:
\be
n_e=\dv{N_e}{E}\,.
\ee
Although Eq.~(\ref{eq:d-function}) and (\ref{eq:d-spectrum}) can be considered an oversimplification, they still allow one to recover some basic properties of IC scattering, especially for the broadband part of the spectra far from either of the cutoff regions. For example, in the Thomson regime (see Eqs.~\eqref{eq:iso_losse_asym} and \eqref{eq:mean_photon_iso_asym}), the energy losses and the mean photon energy depend quadratically on energy: \(\dot{E}\propto E^2\) and \(\bar{\omega}\propto E^2\), respectively. If the electron distribution is a power law, \(n_e\propto E^{-\alpha}\), then Eq.~(\ref{eq:d-spectrum}) yields the standard slope of the Thomson (or synchrotron) spectra \citep[see, e.g.,][]{2011hea..book.....L}:
\be\label{eq:Thomson_slope}
\dot{n}_\gamma\propto \omega^{-(\alpha+1)/2}\,.
\ee
In the Klein-Nishina regime, the energy loss rate is constant, \(\dot{E}\propto \mathrm{const}\) (ignoring the logarithm dependence, see Eq.~\eqref{eq:iso_losse_asym}), and the mean photon energy is \(\bar{\omega}\propto E\), thus for the power-law distribution of electrons one obtains \citep[see, e.g., in][]{1970RvMP...42..237B}
\be\label{eq:KN_slope}
\dot{n}_\gamma\propto \omega^{-(\alpha+1)}\,.
\ee

If the distribution of the target photons is sufficiently broad, then the width target spectrum needs  accounting for. This can also be addressed under the \(\delta\)-function approximation:
\be\label{eq:d-function-photons}
\dot{n}_{\gamma,0}=\int \frac1{\omega}\abs{\dv{\dot{E}}{\ve}}\delta\qty(\omega-\bar{\omega}\qty(E,\ve))\dd{\ve}\,,
\ee
where \(\dd{\dot{E}}\) is the electron energy loss rate caused by the interaction with target photons that have their energy exclusively in the range from \(\ve\) to \(\ve+\dd{\ve}\).

If the distribution of the target photons is sufficiently broad, then at least a fraction of the target photons is up scattered in the Thomson regime. Since the scattering cross-section in the Klein-Nishina regime is smaller than the Thomson cross-section, the part of the spectrum formed in the Thomson regime should reflect the key spectral properties. These features can be studied by setting \(\bar{\omega}\propto \ve E^2\) and  \(\dd{\dot{E}}/\dd{\ve}\propto \ve n\mysub{ph}(\ve) E^2\), where \(n\mysub{ph}\) is the energy distribution of the target photons. Thus, one obtains
  \be\label{eq:thomson_double}
  \dot{n}_\gamma\mysup{T}\propto\int \dd{E}\int\dd{\ve}  \frac1{\omega}E^2\ve n_e(E)n\mysub{ph}(\ve)\delta\qty(\omega-\frac{\ve E^2}{m_e^2c^4})\,.
  \ee

The integration over the \(\delta\)-function helps to clearly reveal the production of the
broadband spectrum in the Thomson limit. Before writing the resulting equation, we note that given the presence of
the \(\delta\)-function term, one does not need to account for the kinematic constraints on the energies of the
interacting particles. However, to ensure that the scattering proceeds in the Thomson regime, we
introduce a Heaviside function that determines the maximum frequency of the scattered photons: \(\Theta\qty(1-\ve E/(m_e^2c^4))\)
(note that in this section we omit some numerical factors, this, however, does not influence the conclusions). Thus, one obtains
  \be\label{eq:thomson_double2}
  \dot{n}_\gamma\mysup{T}\propto\int \dd{E}  \frac1{E^2} n_e(E)n\mysub{ph}\qty(\frac{\omega m_e^2c^4}{E^2})\Theta\qty(E-\omega)=
  \ee
  \[
\int\limits_{\omega}^{\infty} \dd{E}  \frac1{E^2} n_e(E)n\mysub{ph}\qty(\frac{\omega m_e^2c^4}{E^2})\,,    
\]
where the lower energy limit in the integral is due to the Heaviside function, i.e., it is imposed by the Klein-Nishina cutoff, which, according to the assumptions introduced, is equivalent to an obvious requirement, \(E>\omega\).

Let us assume that one deals with a power-law distributions of electrons and target photons:
\be
n_e\propto E^{-\alpha}\Theta\qty(E-E\mysub{min})\Theta\qty(E\mysub{max}-E)
\ee
and 
\be
n\mysub{ph}\propto \ve^{-\beta}\Theta\qty(\ve-\ve\mysub{min})\Theta\qty(\ve\mysub{max}-\ve)\,.
\ee
Provided that  \(E\mysub{max}>m_ec^2\sqrt{\omega/\ve\mysub{max}}\) and \(E\mysub{min}<m_ec^2\sqrt{\omega/\ve\mysub{min}}\), the integral in Eq.~(\ref{eq:thomson_double2}) is a simple power-law function:
\be\label{eq:thomson_double3}
\dot{n}_\gamma\mysup{T}\propto\omega^{-\beta}\int\limits_{\tilde{E}\mysub{min}}^{\tilde{E}\mysub{max}} \dd{E}  E^{2\beta-\alpha-2} \,,
\ee
where the integral limits are determined by the following conditions
\be\label{eq:low_limit}
\tilde{E}\mysub{min}=\max\qty(\omega,E\mysub{min},m_ec^2\sqrt{\frac{\omega}{\ve\mysub{max}}})
\ee
and
\be\label{eq:high_limit}
\tilde{E}\mysub{max}=\min\qty(E\mysub{max},m_ec^2\sqrt{\frac{\omega}{\ve\mysub{min}}})\,.
\ee
Provided \(2\beta-\alpha\neq1\) the final expression is
\be\label{eq:thomson_double4}
\dot{n}_\gamma\mysup{T}\propto\omega^{-\beta}\frac{1}{2\beta-\alpha-1}\qty(\tilde{E}^{2\beta-\alpha-1}\mysub{max}-\tilde{E}^{2\beta-\alpha-1}\mysub{min})\,,
\ee
where the leading term is determined by the sign of the exponent, \(2\beta-\alpha-1\).

\begin{figure}
  \plotone{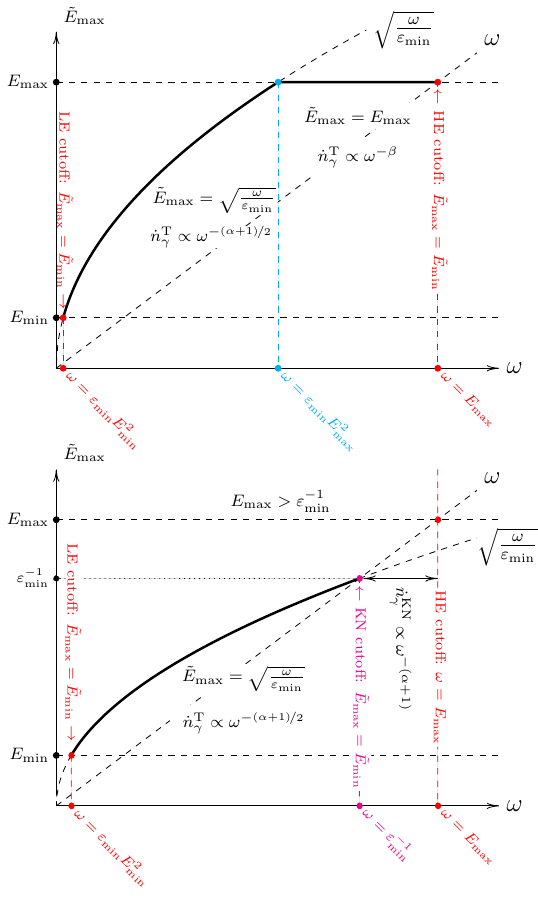}
  \caption{Dependence of \(\tilde{E}\mysub{max}\) from Eq.~(\ref{eq:high_limit}) on the upscattered photon energy together with the conditions that determine the cutoff energy. Note that in the figure labels we omit \(m_ec^2\) factors. \label{fig:upper_bound}}
\end{figure}

If the sign of this exponent is positive, i.e. \(\alpha < 2\beta -1\), then the resulting spectrum is approximately
\be\label{eq:thomson_double5}
\dot{n}_\gamma\mysup{T}\propto\left\{
  \begin{matrix}
    \omega^{-(\alpha+1)/2}&\qq{if} \omega<\frac{\ve\mysub{min}E\mysub{max}^2}{m_e^2c^4}\,,\\
    \omega^{-\beta}&\qq{if} \omega>\frac{\ve\mysub{min}E\mysub{max}^2}{m_e^2c^4}\,.\\
  \end{matrix}
\right.
\ee
Thus, it can be seen that the spectrum keeps the same slope as that predicted by the standard Thomson estimate. If the IC spectrum extends into the energy range where the lower energy part of the spectrum doesn't make any contribution, \(\omega>\ve\mysub{min}E\mysub{max}^2/(m_e^2c^4)\), the IC spectrum is determined by the slope of the target photons, \(\dot{n}_\gamma\mysup{T}\propto\omega^{-\beta}\). The relation between the parameters is graphically shown in Fig.~\ref{fig:upper_bound}.

\begin{figure}
  \plotone{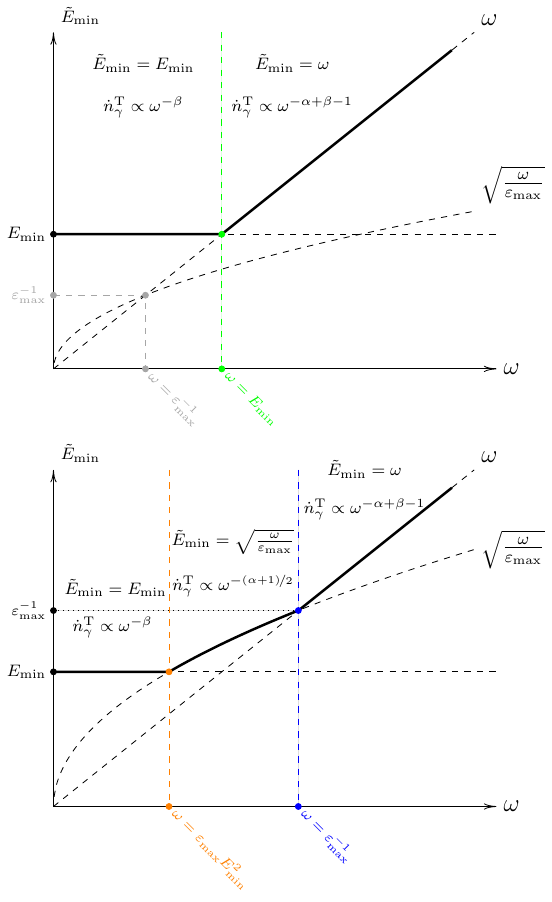}
  \caption{Dependence of \(\tilde{E}\mysub{min}\) from Eq.~(\ref{eq:low_limit}) on the upscattered photon energy. Note that in the figure labels we omit \(m_ec^2\) factors.\label{fig:low_bound}}
\end{figure}

A similar effect defines the slope of the low-energy part of the IC spectrum if \(\alpha > 2\beta -1\). In this case, the high-energy part of the target photon spectrum provides the most important contribution, and the slope of IC emission is inherited from the target photon spectrum if \(\omega<\ve\mysub{max}E\mysub{min}^2/(m_e^2c^4)\) and \(\omega<E\mysub{min}\). The dependence of \(\tilde{E}\mysub{min}\) on \(\omega\) is sketched out in Fig.~\ref{fig:low_bound}. If \(E\mysub{min}>m_e^2c^4/\ve\mysub{max}\) the IC spectrum directly translates from \(\omega^{-\beta}\) to \(\omega^{-\alpha+\beta-1}\) at \(\omega=E\mysub{min}\). If \(E\mysub{min}<m_e^2c^4/\ve\mysub{max}\) the transition between these two regimes proceeds through \(\omega^{-(\alpha+1)/2}\), which is realized for \(\ve\mysub{max}E\mysub{min}^2/(m_e^2c^4)<\omega<m_e^2c^4/\ve\mysub{max}\). These spectral properties are summarized by the following expressions (see also Fig.~\ref{fig:low_bound}): 
\be\label{eq:thomson_double6}
\dot{n}_\gamma\mysup{T}\propto\left\{
  \begin{matrix}
    \omega^{-\beta}&\qq{if} &\omega<\frac{\ve\mysub{max}E\mysub{min}^2}{m_e^2c^4}&\qq{and}& \omega<E\mysub{min}\,,\\
    \omega^{-(\alpha+1)/2}&\qq{if}& \omega>\frac{\ve\mysub{max}E\mysub{min}^2}{m_e^2c^4}&\qq{and}&\omega<\frac{m_e^2c^4}{\ve\mysub{max}}\,,\\
    \omega^{-(\alpha+1)+\beta}&\qq{if}& \omega>E\mysub{min}&\qq{and}&\omega>\frac{m_e^2c^4}{\ve\mysub{max}}\,. 
  \end{matrix}
\right.
\ee


\begin{figure}
  \plotone{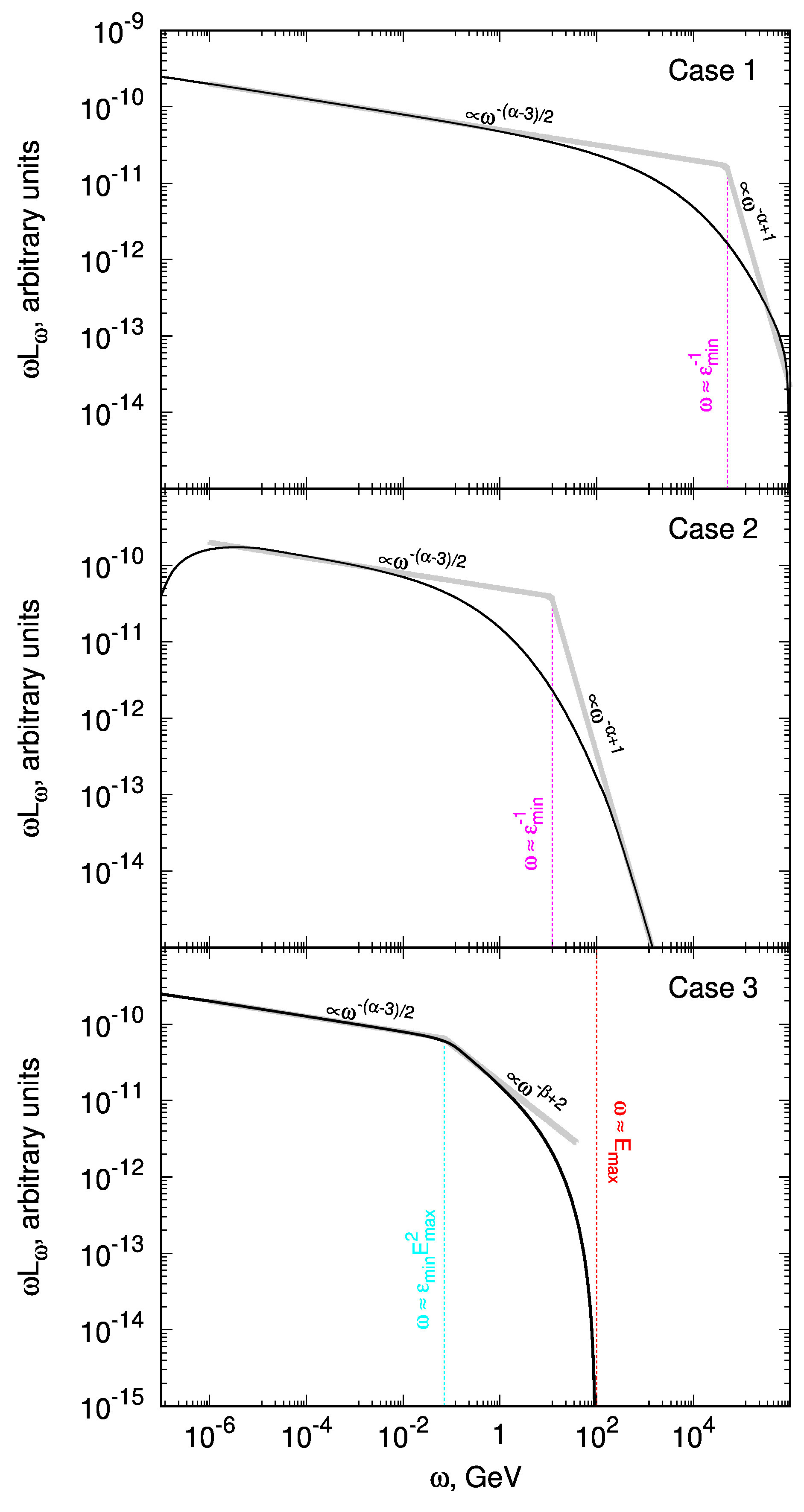}
  \caption{Numerical computation of IC spectrum produced on a power-law target photons with \(\beta=2.5\). Upper panel (``Case 1''): minimum and maximum energies of target photons are \(\ve\mysub{min}=10^{-3}\unit{eV}\)  and \(\ve\mysub{max}=1\unit{keV}\), respectively; the electron maximum energy was set to \(E\mysub{max}=1\unit{PeV}\). Middle panel (``Case 2''): minimum and maximum energies of target photons are \(\ve\mysub{min}=10\unit{eV}\)  and \(\ve\mysub{max}=3\unit{keV}\), respectively; the electron maximum energy was set to \(E\mysub{max}=1\unit{PeV}\). Bottom panel (``Case 3''): minimum and maximum energies of target photons are \(\ve\mysub{min}=10^{-3}\unit{eV}\)  and \(\ve\mysub{max}=1\unit{keV}\), respectively; the electron maximum energy was set to \(E\mysub{max}=100\unit{GeV}\). The electron energy distribution was assumed to be a power law with \(\alpha=3.2\) above \(E\mysub{min}=1\unit{MeV}\). The solid guide lines indicate the analytic slopes expected from Eq.~(\ref{eq:thomson_double5}) and (\ref{eq:KN_slope}) (in the Klein-Nishina limit) and the dashed guide lines indicate the positions of spectral transformations given by Eqs.~(\ref{eq:thomson_double5}) and (\ref{eq:kn_condition}). The slope labels show the energy flux spectral indices. \label{steep_photons}}
\end{figure}

\begin{figure}
  \plotone{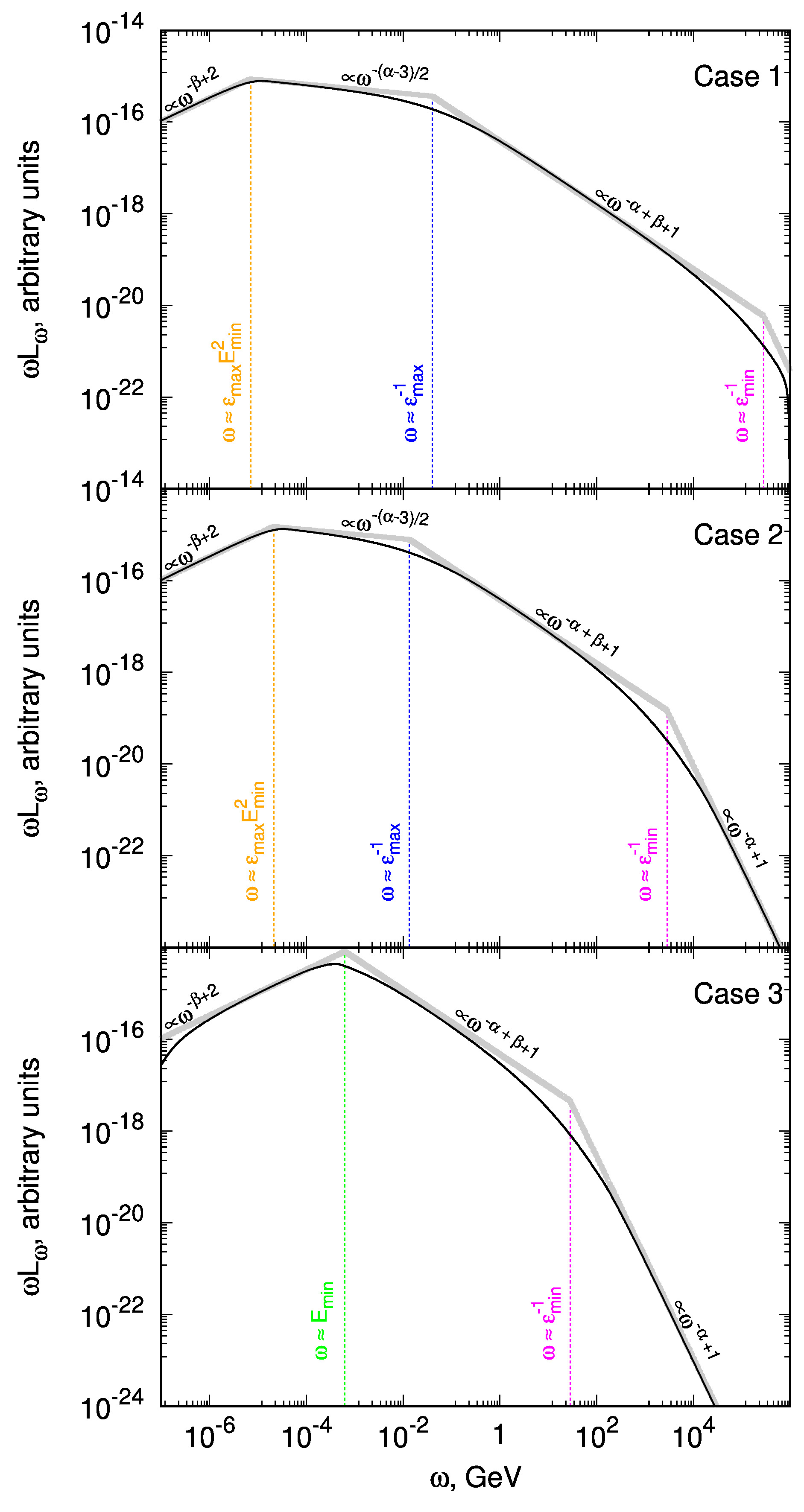}
  \caption{Numerical computation of IC spectrum produced on a power-law target photons with \(\beta=1.5\). Upper panel (``Case 1''): minimum and maximum energies of target photons are \(\ve\mysub{min}=10^{-3}\unit{eV}\)  and \(\ve\mysub{max}=1\unit{keV}\), respectively. Middle panel (``Case 2''): minimum and maximum energies of target photons are \(\ve\mysub{min}=10^{-1}\unit{eV}\)  and \(\ve\mysub{max}=3\unit{keV}\), respectively. Bottom panel (``Case 3''): minimum and maximum energies of target photons are \(\ve\mysub{min}=10\unit{eV}\)  and \(\ve\mysub{max}=100\unit{keV}\), respectively.  The electron energy distribution was assumed to be a power law with \(\alpha=3.2\) between \(E\mysub{min}=1\unit{MeV}\) and \(E\mysub{max}=1\unit{PeV}\). The solid guide lines shown are the analytic slopes expected from Eqs.~(\ref{eq:thomson_double6}) and (\ref{eq:KN_slope}) (in the Klein-Nishina limit) and dashed guide lines indicate the positions of spectral transformations given by Eqs.~(\ref{eq:thomson_double6}) and (\ref{eq:kn_condition}). The slope labels show the energy flux spectral indices.\label{hard_photons}}
\end{figure}

The resultant IC spectra obtained by numerical integration of the differential cross-section over power-law distributions of target photons and electrons are shown in Figs.~\ref{steep_photons} and \ref{hard_photons}. The simple analytic dependencies shown in the figures are given by Eqs.~(\ref{eq:thomson_double5}) and (\ref{eq:thomson_double6}). Also, it can be seen from Fig.~\ref{hard_photons} that  Eq.~(\ref{eq:KN_slope}) describes the spectral slope in the part of the spectrum generated in the Klein-Nishina regime.

Another important question is in which energy interval the revealed power-law dependencies are relevant. The approach used is relevant if the integration interval is sufficiently broad. Thus, the condition of the applicability of the obtained results is 
\be
\tilde{E}\mysub{min}\ll\tilde{E}\mysub{max}\,.
\ee
Once the integral limits approach each other, the power-law behavior becomes distorted, and the integral in Eq.~(\ref{eq:thomson_double3}) starts to vanish, i.e., the condition \(\tilde{E}\mysub{min}\sim\tilde{E}\mysub{max}\) defines the position of the spectral cutoffs. 
The low-energy cutoff is then simply given by the condition
\be
\omega\mysub{min}\approx \frac{\ve\mysub{min}E\mysub{min}^2}{m_e^2c^4}\,.
\ee
For the high-energy cutoff, the determination of the conditions is a little more involved. 
Unless the parameters are tuned, it is natural to expect that in the high \(\omega\) regime the low energy limit of the integral, Eq.~(\ref{eq:low_limit}),  is simply \(\omega\). The upper bound can be either \(E\mysub{max}\) or \(m_ec^2\sqrt{\frac{\omega}{\ve\mysub{min}}}\) (see in Fig.~\ref{fig:upper_bound} for a sketch). In the former case, the disappearance of the integral is caused by \(\omega\rightarrow E\mysub{max}\) and the spectrum should completely disappear at \(E\mysub{max}\).
This is shown by the IC spectrum computed for \(E\mysub{max}=100\unit{GeV}\) in the bottom panel of Fig.~\ref{steep_photons}.

In the case when \(E\mysub{max}>m_e^2c^4/\ve\mysub{min}\), one should expect a transition to the Klein-Nishina regime when \(\omega\rightarrow m_ec^2\sqrt{\frac{\omega}{\ve\mysub{min}}}\). This means that at the gamma-ray energy
\be\label{eq:kn_condition}
\omega\mysub{kn}\sim\frac{m_e^2c^4}{\ve\mysub{min}}\,,
\ee
the IC spectrum should obtain a typical slope for the Klein-Nishina regime: \(\dot{n}_\gamma\propto \omega^{-(\alpha+1)}\)  (see in Fig.~\ref{fig:upper_bound}).
This transformation is illustrated by curves computed for  \(\ve\mysub{min}=10\unit{eV}\) in Figs.~\ref{steep_photons}~and~\ref{hard_photons}.

However, we note that the IC spectra can appear significantly harder than that expected to be produced via interactions in the Klein-Nishina regime, if the distribution of target photons extends to sufficiently low energies (i.e., the condition given by Eq.~\eqref{eq:kn_condition} is not fulfilled --- see the spectra computed for \(\ve\mysub{min}=10^{-3}\unit{eV}\) in the top panels of Figs.~\ref{steep_photons} and \ref{hard_photons}).

\begin{figure}
  \plotone{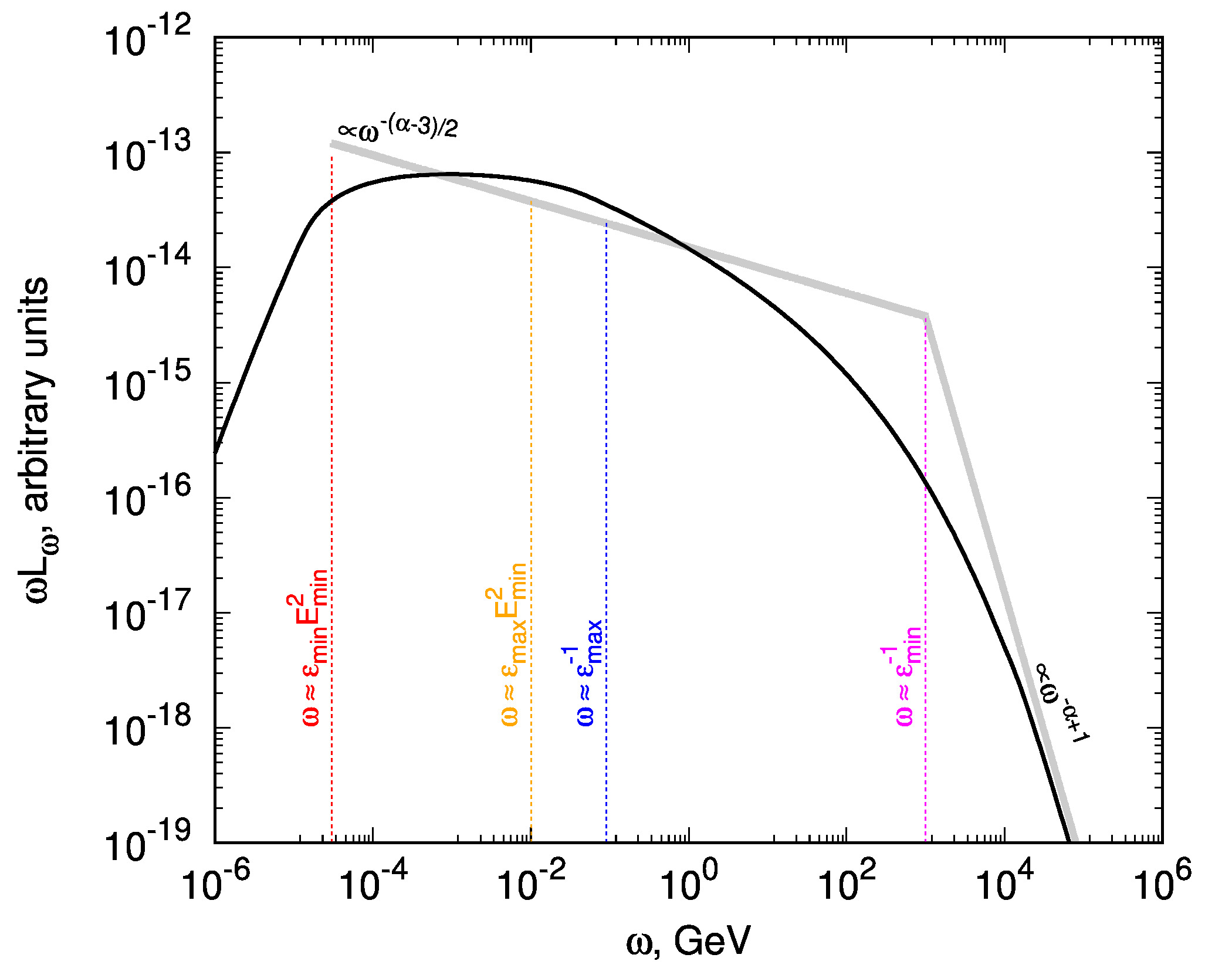}
  \caption{Numerical computation of IC spectrum produced on a power-law target photons with \(\beta=2.2\) between \(\ve\mysub{min}=0.1\unit{eV}\)  and \(\ve\mysub{max}=1\unit{keV}\). The electron energy distribution was assumed to be a power law with \(\alpha=3.4\) between \(E\mysub{min}=100\unit{MeV}\) and \(E\mysub{max}=1\unit{PeV}\). { Guide lines showing the} analytic slopes expected in the Thomson and in the Klein-Nishina regimes; the positions of low- and high-energy spectral cutoffs together with ``logarithmic'' spectral transformations are indicate with labels. The slope labels show the energy flux spectral indices.\label{log_photons}}
\end{figure}

So far, we have assumed that \(\beta\neq (\alpha+1)/2\). However, this specific case deserves special mention as synchrotron emission produced by electrons having a power-law energy distribution gives rise to a power-law spectrum with photon index \((\alpha+1)/2\).  In this case, all the revealed photon indexes of the IC component generated in the Thomson regime correspond to the same spectral slope, as one has
\be
\frac{\alpha+1}{2}=\eval{\beta}_{\beta=(\alpha+1)/2}^{} = \eval{\alpha+1 - \beta}_{\beta=(\alpha+1)/2}^{}\,.
\ee
Therefore, the discussed spectral breaks vanish and the Thomson spectrum becomes a single power-law component with the standard Thomson photon index \((\alpha+1)/2\). This consideration neglects the influence of the logarithmic factors. Accounting for these factors, one obtains
\be\label{eq:thomson_double_log}
\dot{n}_\gamma\mysup{T}\propto\omega^{-(\alpha+1)/2}\log\qty(\frac{\tilde{E}\mysub{max}}{\tilde{E}\mysub{min}})\,.
\ee
The function under the logarithm experience breaks at each of the frequencies defined by Eqs.~\eqref{eq:thomson_double5}~and~\eqref{eq:thomson_double6}, which causes deviations from the precise power-law dependence even if the emission is formed entirely in the Thomson regime, as shown in Fig.~\ref{log_photons}.

\section{Discussion}\label{sec:sum}

While the spectral properties of the IC component generated by interactions on a target that is characterized by a
  narrow energy distribution are well understood, such results cannot be a priori generalized to cases in which the relativistic electrons interact with a broad, i.e., a wide spread of frequencies over several orders of magnitude,
  distribution of photons. To reveal the ``anatomy'' of the IC components generated under such conditions, we
have analyzed the IC scattering process on a power-law distribution of target photons analytically under the
\(\delta\)-function approximation, verifying these findings through comparisons with the numerical integration
results. We demonstrate that the generated IC component has a broken-power-law shape, where, for certain parameter
  combinations, we may expect up to three spectral breaks. The obtained spectral slopes depend on the power-law index of
  the electron distribution, \(\alpha\), and on the photon index of the target photons' distribution, \(\beta\).
Figures~\ref{steep_photons} and \ref{hard_photons} show that the determined analytical properties adequately reproduce
the spectral properties obtained using accurate numerical calculations.

The obtained results reveal the key factors determining the spectral transitions. Under the \(\delta\)-function
approximation, these factors are reduced to the dependence of the integration limits on the gamma-ray energy, which
are summarized in Figs.~\ref{fig:upper_bound}~and~\ref{fig:low_bound}.

The analysis performed involves some simplifying assumptions. In particular, it was assumed that the target photons have a power-law distribution in the entire frequency range, between \(\ve\mysub{min}\) and \(\ve\mysub{max}\). At high frequencies, one often expects a relatively sharp cutoff, which can be reasonably approximated by a truncated power-law distribution. In contrast, at the lower frequency end, one often expects a presence of a cooling break, which may cause a spectral transformation.

The analysis of the influence of a cooling break (say at \(E\mysub{br}\)) is straightforward given that the electron distribution can be represented as
\be\label{eq:bpl_electrons}
\begin{split}
n_{e}\propto\quad& E^{-(\alpha-1)}\Theta\qty(E-E\mysub{min})\Theta\qty(E\mysub{br}-E)+
\\
 & E\mysub{br}E^{-\alpha}\Theta\qty(E-E\mysub{br})\Theta\qty(E\mysub{max}-E)\,.
\end{split}
\ee
If one assumes that the target photons have a synchrotron origin, the photon energy distribution should also be a broken power-law spectrum:
\be\label{eq:bpl_target}
\begin{split}
n\mysub{ph}\propto\quad& \ve^{\nicefrac{-\alpha}{2}}\Theta\qty(\ve-\ve\mysub{min})\Theta\qty(\ve\mysub{br}-\ve)+\\
                  &\ve\mysub{br}^{\nicefrac{1}{2}}\ve^{\nicefrac{-(\alpha+1)}{2}}\Theta\qty(\ve-\ve\mysub{br})\Theta\qty(\ve\mysub{max}-\ve)\,,
\end{split}
\ee
where the break and cutoff positions should satisfy the following relations: \(\ve\mysub{min} = \ve\mysub{max}\qty(E\mysub{min}/E\mysub{max})^2\) and \(\ve\mysub{br} = \ve\mysub{max}\qty(E\mysub{br}/E\mysub{max})^2\).

Since Eq.~\eqref{eq:thomson_double2} has a linear dependence on the densities of the target photons and electrons, the resulting gamma-ray spectrum should be a linear superposition of spectra computed for each of the terms in Eqs.~\eqref{eq:bpl_target} and \eqref{eq:bpl_electrons}. The final spectrum is therefore a superposition of four components, each of which is produced by a power-law energy distribution of electrons on a power-law energy distribution of target photons, and thus can be obtained in the framework of the suggested approach.

Another important feature of the IC component is the transition from the Thomson to the Klein-Nishina regimes. This
transformation leads to a significant spectral softening that may strongly affect the properties of the very-high-energy (VHE) emission. We
have shown that if the target photon spectrum extends to sufficiently low energies,
\(\ve\mysub{min}<m_e^2c^4/E\mysub{max}\), then the entire IC component should be dominated by photons generated in the Thomson
regime, thus the Klein-Nishina effect would not cause any strong spectral softening.

\begin{figure}
  \plotone{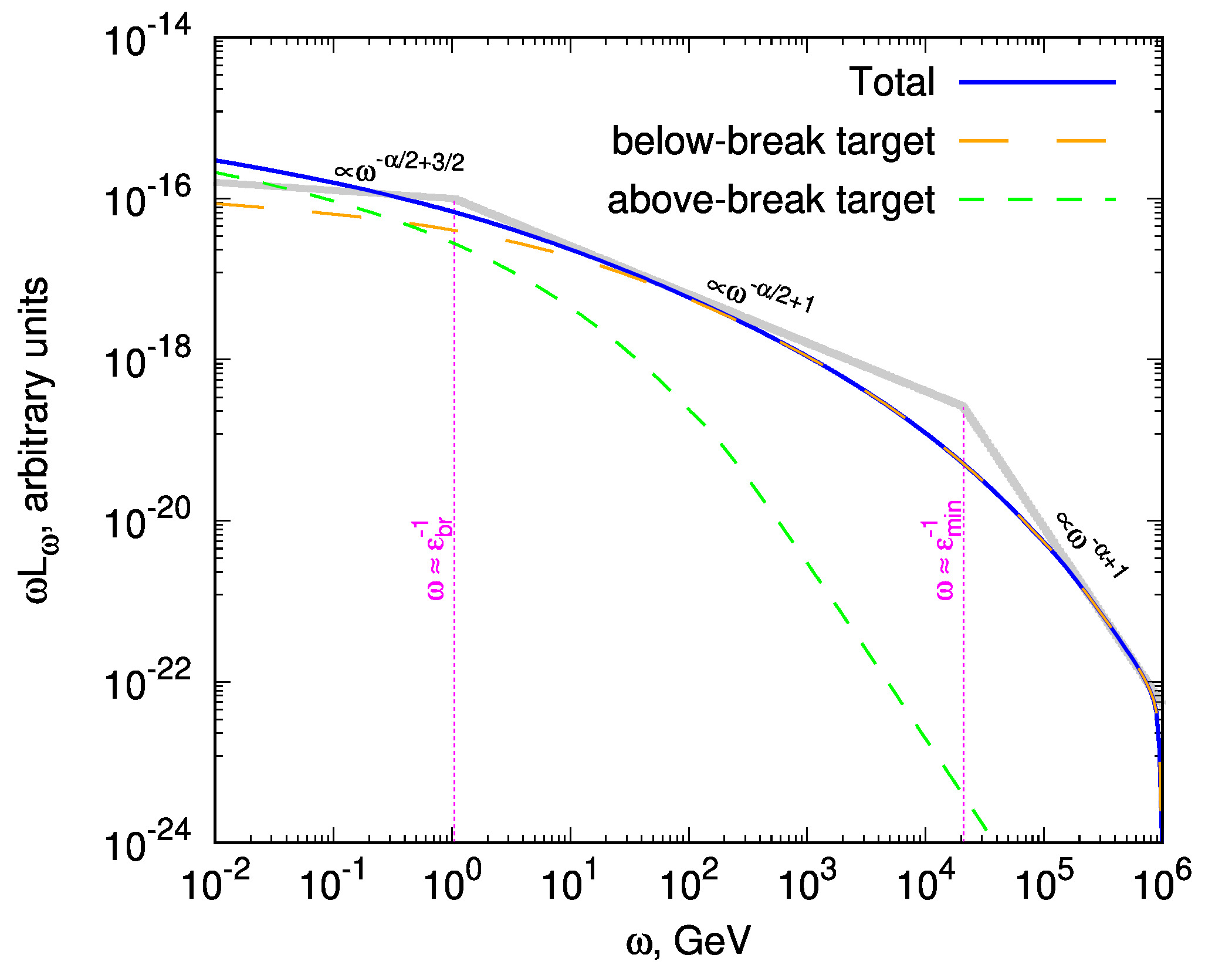}
  \caption{Numerical computation of IC spectrum produced on a broken-power-law target photons. The photon indexes were adopted \(\beta_1=1.6\) and \(\beta_1=2.1\)  from \(\ve\mysub{min}=10^{-2}\unit{eV}\)  to \(\ve\mysub{br}=10\unit{eV}\) and from \(\ve\mysub{br}=10\unit{eV}\) to \(\ve\mysub{max}=1\unit{keV}\), respectively. The electron energy distribution was assumed to be a power law with \(\alpha=3.2\) between \(E\mysub{min}=1\unit{MeV}\) and \(E\mysub{max}=1\unit{PeV}\). { Guide lines are provided to indicate} the analytic slopes expected in the high-energy part of the IC spectrum.\label{cutoff}}
\end{figure}

To illustrate the influence of the Klein-Nishina effect, we qualitatively describe the spectral transformations
expected from the above analysis for the case in which the electron distribution features a cooling break { and the target photons are provided by the synchrotron mechanism}. As we are focused on the high-energy part of the spectrum, we can approximate
the electron distribution as a single power-law spectrum: \(n_e\propto E^{-\alpha}\). As the low-energy photons define
the position of the transition to the Klein-Nishina dominated spectrum, the influence of the cooling break is important
for the target photons. Thus, to describe the photon distribution, we use Eq.~\eqref{eq:bpl_target}. When the electrons
up-scatter the high-energy part of the photon target, a component with photon index \((\alpha+1)/2\) is formed (see
Fig.~\ref{log_photons} for an example, but note that one should replace \(\ve\mysub{min}\) with \(\ve\mysub{br}\)). This
component extends to an energy \(m_e^2c^4/\ve\mysub{br}\), if the cooling break is formed at a sufficiently high energy,
\(\ve\mysub{br}>m_e^2c^4/E\mysub{max}\). Above this energy, \(m_e^2c^4/\ve\mysub{br}\), there is a contribution from the
low-energy part of the photon target, which is still up-scattered in the Thomson regime. Equation (\ref{eq:thomson_double6})
defines the spectral properties in this case, thus the spectrum slope is \(\alpha+1-\beta=(\alpha+1)/2 + 1/2\) (provided
that \(\beta=\alpha/2\)). Thus, at energy \(m_e^2c^4/\ve\mysub{br}\) one expects a break in the spectral index by \(1/2\). At higher energies,
power-law spectrum should extend up to \(m_e^2c^4/\ve\mysub{min}\). Finally, if \(E\mysub{max}>m_e^2c^4/\ve\mysub{min}\), the IC emission
should obtain the typical Klein-Nishina photon index of \(\alpha+1\). This qualitative description is compared with the numerical simulations in Fig.~\ref{cutoff} from which one can see a reasonable agreement with the numerical simulations.

A broad distribution of the target photons in astrophysical sources can be formed by a superposition of blackbody components with different temperatures, e.g., from a multi-color accretion disk, or when the target photons are provided by a non-thermal radiation mechanism. { Our results are applicable, with a reasonable level of accuracy, in both of these cases. However in the latter case, the target photons often offer a particularly broad power-law energy distribution over many decades in frequency}. The effects discussed above therefore reveal themselves { most} clearly in this case. One of the most common scenarios, when a non-thermal mechanism provides target photons for IC scattering, is the so-called synchrotron self-Compton (SSC) mechanism, in which the target photons are supplied by synchrotron emission. In particular, the SSC process provides the most natural scenario for interpreting the gamma-ray emission from sources such as AGN jets \citep[e.g.,][]{1992ApJ...397L...5M} and GRBs \citep[see][and references therein]{2016MNRAS.460.2036D}. In the case of the idealized SSC scenario, that is, the electrons obey a single power-law component, and exclusively up-scatter synchrotron photons, then some of the revealed spectral breaks vanish and the part of the IC component formed in the Thomson regime obtains the standard slope, \((\alpha+1)/2\).  There are, however, logarithmic terms that can cause a considerable deformation of this power-law behavior, as shown in Fig.~\ref{log_photons}.

{To make more concrete this discussion, we finish with an application of our results to a particular GRB data set. The last few years has seen exciting results in the domain of gamma-ray astronomy, with the detection of TeV gamma-ray emission from GRB afterglows} \citep{2019Natur.575..464A,2019Natur.575..455M,2021Sci...372.1081H,doi:10.1126/science.adg9328}. {For the most local of these events, GRB190829, the proximity of the GRB allowed the accurate determination} of the spectral properties of the VHE emission over a broad energy range, spanning from a few hundred GeV to several TeV. The H.E.S.S. observations of GRB190829A revealed that the intrinsic\footnote{I.e., corrected for the attenuation on the extragalactic background light.} gamma-ray spectrum between \(180\unit{GeV}\) and \(3.3\unit{TeV}\), measured approximately 5 hours after the trigger, {showed no evident signs of the Klein-Nishina softening.  Indeed, the spectral index of the TeV component was measured to be \(\gamma\mysub{VHE}=2.06\pm0.36\) (here we sum up both statistical and systematic uncertainties), which agrees well with the X-ray spectral index measured at the same epoch by the Swift X-ray telescope, \(\gamma\mysub{xrt}=2.03\pm0.06\) \citep{2021Sci...372.1081H}. The almost identical values for the X-ray and TeV spectral slopes suggest that the IC scattering proceeds in the Thomson regime. Even if one accounts for the uncertainties, the difference between the spectral slopes is smaller than \(0.5\)}. According to the results {summarised in sect.~\ref{sec:sum}}, the matching X-ray and VHE slopes imply that the cooling break in the synchrotron spectrum should be at sufficiently low frequencies. Since the production region moves relativistically, say with a bulk Lorentz factor \(\Gamma\), the cooled synchrotron spectrum should extend below {an energy of}
  \be
  \ve\mysub{br} < \frac{\Gamma^2m_e^2c^4}{\omega\mysub{vhe}}\approx 0.1 \Gamma^{2}\qty(\frac{\omega\mysub{vhe}}{3\unit{TeV}})^{-1}\unit{eV}\,.
  \ee
  (Note that here the photon and gamma-ray energies are in the observer frame.) 

  On the other hand, the optical {observation results suggest} that the cooling break in the
  target photons appears at higher frequencies, \(\ve\mysub{br}>100\unit{eV}\) \citep[and, e.g., Fig.~S5 in \citealt{2021Sci...372.1081H}]{2021A&A...646A..50H}. Thus, the minimum Lorentz factor that allows
  to explain simultaneously the X-ray and VHE data detected from GRB190829A, is \(\Gamma\mysub{min}\approx
  30\). However, the self-similar solution for a relativistic blast wave \citep{1976PhFl...19.1130B} predicts for the stage of afterglow at \(t\sim5\unit{h}\) a significantly smaller bulk Lorentz factor, \(\Gamma\lesssim5\), for the conditions expected at the explosion of 
  progenitor star of GRB190829A \citep[e.g.,][]{2021Sci...372.1081H}.

  The simple analysis above allows one to reveal the limitations of one zone SSC models in explaining H.E.S.S. observations of GRB190829A. We note that a similar conclusion was obtained with Markov Chain Monte Carlo study of parameter space in the SSC scenario  for this GRB \citep{2021Sci...372.1081H}.

\begin{acknowledgments}
{The authors thank anonymous referee and V.~Bosch-Ramon for useful comments and suggestions.} DK acknowledges the support of RSF grant No. 21-12-00416. {AT acknowledges support from DESY (Zeuthen, Germany), a member of the Helmholtz Association HGF.}
\end{acknowledgments}

%

\vspace{5mm}










\end{document}